\documentclass[11pt]{article}
\usepackage{amsmath,amssymb,amsthm,amsxtra,overpic,bbm,bm,epsfig,ulem}
\usepackage{color}
\def\thefootnote{\fnsymbol{footnote}}

\begin{document}

\vspace{0.2cm}

\begin{center}
{\Large\bf Geometry of the effective Majorana neutrino mass
in the $0\nu\beta\beta$ decay}
\end{center}

\vspace{0.2cm}

\begin{center}
{\bf Zhi-zhong Xing $^{a,b}$} \footnote{E-mail: xingzz@ihep.ac.cn}
\quad {\bf Ye-Ling Zhou $^{a}$} \footnote{E-mail:
zhouyeling@ihep.ac.cn}
\\
{$^a$Institute of High Energy Physics, Chinese Academy of
Sciences, P.O. Box 918, Beijing 100049, China \\
$^b$Center for High Energy Physics, Peking University, Beijing
100080, China}
\end{center}

\vspace{1.5cm}
\begin{abstract}
The neutrinoless double-beta ($0\nu\beta\beta$) decay is a unique
process to identify the Majorana nature of massive neutrinos, and
its rate depends on the size of the effective Majorana neutrino mass
$\langle m\rangle^{}_{ee}$. We put forward a novel ``coupling-rod"
diagram to describe $\langle m\rangle^{}_{ee}$ in the complex plane,
by which the effects of the neutrino mass ordering and CP-violating
phases on $\langle m\rangle^{}_{ee}$ are intuitively understood. We
show that this geometric language allows us to easily obtain the
maximum and minimum of $|\langle m\rangle^{}_{ee}|$. It remains
usable even if there is a kind of new physics contributing to
$\langle m\rangle^{}_{ee}$, and it can also be extended to describe
the effective Majorana masses $\langle m\rangle^{}_{e\mu}$, $\langle
m\rangle^{}_{e\tau}$, $\langle m\rangle^{}_{\mu\mu}$, $\langle
m\rangle^{}_{\mu\tau}$ and $\langle m\rangle^{}_{\tau\tau}$ which
may appear in some other lepton-number-violating processes.
\end{abstract}

\begin{flushleft}
\hspace{0.8cm} PACS number(s): 14.60.Pq, 13.15.+g, 25.30.Pt
\end{flushleft}

\def\thefootnote{\arabic{footnote}}
\setcounter{footnote}{0}

\newpage

\section{Introduction}

Whether massive neutrinos are the Majorana particles remains an open
question in particle physics. By definition, a Majorana neutrino is
its own antiparticle \cite{Majorana}, and this consequently leads to
lepton number violation. Because the masses of three known neutrinos
are extremely small, the only feasible way to identify their
Majorana nature is to detect the neutrinoless double-beta
($0\nu\beta\beta$) decay of some even-even nuclei \cite{Furry}:
$N(A, Z) \to N(A, Z+2) + 2 e^-$, in which the lepton number is
violated by two units. In the basis where flavor and mass
eigenstates of the charged leptons coincide with each other, the
$0\nu\beta\beta$ decay rate is controlled by the $(e,e)$ element of
the effective Majorana neutrino mass matrix \cite{2B}
\begin{eqnarray}
M^{}_\nu = \left(\begin{matrix} \langle m\rangle^{}_{ee} &
\langle m\rangle^{}_{e\mu} & \langle m\rangle^{}_{e\tau} \cr
\langle m\rangle^{}_{\mu e} & \langle m\rangle^{}_{\mu\mu} &
\langle m\rangle^{}_{\mu\tau} \cr
\langle m\rangle^{}_{\tau e} & \langle m\rangle^{}_{\tau\mu} &
\langle m\rangle^{}_{\tau\tau} \cr \end{matrix} \right) \; ,
\hspace{1.2cm}
\langle m\rangle^{}_{\alpha\beta} \equiv \sum_i \left(m^{}_i
U^{}_{\alpha i} U^{}_{\beta i} \right) \; ,
\end{eqnarray}
where the Greek subscripts run over $e$, $\mu$ and $\tau$, $m^{}_i$
denotes the $i$-th neutrino mass, and $U^{}_{\alpha i}$ stands for
the corresponding element of the lepton flavor mixing matrix $U$
\cite{MNSP}. Of course, $\langle m\rangle^{}_{\alpha\beta} = \langle
m\rangle^{}_{\beta\alpha}$ holds for symmetric $M^{}_\nu$. In the
standard three-flavor scheme, $U$ is a unitary matrix and can
therefore be parametrized in terms of three flavor mixing angles and
three CP-violating phases:
\begin{eqnarray}
U = \left( \begin{matrix}
c^{}_{12} c^{}_{13} & s^{}_{12} c^{}_{13} & s^{}_{13} e^{-{\rm i}
\delta} \cr -s^{}_{12} c^{}_{23} - c^{}_{12} s^{}_{13} s^{}_{23}
e^{{\rm i} \delta} & c^{}_{12} c^{}_{23} - s^{}_{12} s^{}_{13}
s^{}_{23} e^{{\rm i} \delta} & c^{}_{13} s^{}_{23} \cr s^{}_{12}
s^{}_{23} - c^{}_{12} s^{}_{13} c^{}_{23} e^{{\rm i} \delta} &
-c^{}_{12} s^{}_{23} - s^{}_{12} s^{}_{13} c^{}_{23} e^{{\rm i}
\delta} & c^{}_{13} c^{}_{23} \cr
\end{matrix} \right) P^{}_\nu \; ,
\end{eqnarray}
where $c^{}_{ij} \equiv \cos\theta^{}_{ij}$, $s^{}_{ij} \equiv
\sin\theta^{}_{ij}$ (for $ij = 12, 13, 23$), and $P^{}_\nu = {\rm
Diag}\left\{e^{{\rm i}\rho/2}, 1, e^{{\rm i} (\delta
+\sigma/2)}\right\}$. As a result, the effective mass term of the
$0\nu\beta\beta$ decay reads
\begin{eqnarray}
\langle m\rangle^{}_{ee} = m^{}_1 |U^{}_{e 1}|^2 e^{{\rm i}\rho}
+ m^{}_2 |U^{}_{e2}|^2 + m^{}_3 |U^{}_{e3}|^2 e^{{\rm i} \sigma} =
m^{}_1 c^2_{12} c^2_{13} e^{{\rm i}\rho} + m^{}_2 s^2_{12} c^2_{13}
+ m^{}_3 s^2_{13} e^{{\rm i}\sigma} \; .
\end{eqnarray}
So far the values of $\theta^{}_{12}$, $\theta^{}_{13}$ and
$\theta^{}_{23}$ have been determined to a good degree of accuracy
from current neutrino oscillation data, but the three phase
parameters remain unknown \cite{PDG}. While the value of $\Delta
m^2_{21} \equiv m^2_2 - m^2_1$ and the absolute value of $\Delta
m^2_{31} \equiv m^2_3 - m^2_1$ are also measured, the sign of
$\Delta m^2_{31}$ and the absolute neutrino mass scale remain
unknown. Hence the size of $\langle m\rangle^{}_{ee}$ suffers from
three kinds of uncertainties even without any new physics pollution:
\begin{itemize}
\item     The unknown absolute neutrino mass scale (i.e., the value of
$m^{}_1$, $m^{}_2$ or $m^{}_3$);
\item     The unknown neutrino mass ordering (i.e., either
$\Delta m^2_{31} >0$ or $\Delta m^2_{31} <0$);
\item     The unknown Majorana CP-violating phases $\rho$ and
$\sigma$ appearing in $|\langle m\rangle^{}_{ee}|$.
\end{itemize}
Up to now, a lot of phenomenological efforts have been made to
probe the parameter space of $\langle m\rangle^{}_{ee}$ and
discuss its sensitivity to possible new physics \cite{Rode}.

In the present work we are going to put forward a novel
``coupling-rod" diagram to describe the salient features of $\langle
m\rangle^{}_{ee}$ in the complex plane, by which the effects of the
neutrino mass ordering and CP-violating phases on $\langle
m\rangle^{}_{ee}$ can be intuitively understood. Some special but
interesting cases, including the behavior of $\langle
m\rangle^{}_{ee}$ with $m^{}_1 =0$ or $m^{}_3 =0$ and the maximum or
minimum of $|\langle m\rangle^{}_{ee}|$ in two different neutrino
mass spectra, are easily explained in this geometric language. We
point out that the coupling-rod diagram remains applicable even if a
kind of new physics, such as an extra light but sterile neutrino,
contributes to $\langle m\rangle^{}_{ee}$. It can also be extended
to provide a vivid description of the effective Majorana neutrino
masses $\langle m\rangle^{}_{e\mu}$, $\langle m\rangle^{}_{e\tau}$,
$\langle m\rangle^{}_{\mu\mu}$, $\langle m\rangle^{}_{\mu\tau}$ and
$\langle m\rangle^{}_{\tau\tau}$, which may show up in
neutrino-antineutrino oscillations and some other
lepton-number-violating processes.

\section{The coupling-rod diagram of $\langle m\rangle^{}_{ee}$}

Given $\Delta m^2_{21} >0$ as established from the solar neutrino
oscillation data, the unfixed sign of $\Delta m^2_{31}$ implies that
the neutrino mass ordering can be either normal (i.e., $m^{}_1 <
m^{}_2 < m^{}_3$) or inverted (i.e., $m^{}_3 < m^{}_1 < m^{}_2$). In
particular, the possibility of $m^{}_1 =0$ or $m^{}_3 =0$ is still
allowed by current experimental data. Because of $m^{}_2 >0$,
together with
\begin{eqnarray}
m^{}_1 \hspace{-0.2cm} & = & \hspace{-0.2cm}
\sqrt{m^2_2 - \Delta m^2_{21}} \; ,
\nonumber \\
m^{}_3 \hspace{-0.2cm} & = & \hspace{-0.2cm}
\sqrt{m^2_2 - \Delta m^2_{21} + \Delta m^2_{31}} \; ,
\end{eqnarray}
we find that it is most convenient to take the nonzero $m^{}_2
U^2_{e 2}$ term as the base vector to geometrically describe
$\langle m\rangle^{}_{ee}$ in the complex plane
\footnote{Some authors have chosen the $m^{}_1 U^2_{e1}$ term as the
base vector to illustrate the geometry of $\langle m\rangle^{}_{ee}$
\cite{Smirnov}. Such a choice has a remarkable disadvantage, because
the $m^{}_1 \to 0$ limit will make this geometric language
invalid.}.
Taking account of the phase convention of $P^{}_\nu$ in Eq. (3),
which allows $U^{}_{e2}$ to be real and positive, we have
\begin{eqnarray}
\overrightarrow{OA} \hspace{-0.2cm} & \equiv & \hspace{-0.2cm}
m^{}_2 U^2_{e2} = m^{}_2 |U^{}_{e2}|^2 \; , \nonumber \\
\overrightarrow{AB} \hspace{-0.2cm} & \equiv & \hspace{-0.2cm}
m^{}_1 U^2_{e1} = m^{}_1 |U^{}_{e1}|^2 e^{{\rm i}\rho} \; , \nonumber \\
\overrightarrow{CO} \hspace{-0.2cm} & \equiv & \hspace{-0.2cm}
m^{}_3 U^2_{e3} = m^{}_3 |U^{}_{e3}|^2 e^{{\rm i}\sigma} \; , \hspace{0.2cm}
\end{eqnarray}
as illustrated in Fig. 1. So the vector $\overrightarrow{CB} =
\overrightarrow{OA} + \overrightarrow{AB} + \overrightarrow{CO}$,
which connects the two circles formed by the rotations of
$\overrightarrow{AB}$ and $\overrightarrow{OC}$ about their
respective origins $A$ and $O$, looks like the ``coupling rod" of a
locomotive and stands for $\langle m\rangle^{}_{ee}$. Depending on
the length of $\overrightarrow{OA}$ and the radii of $\bigodot O$
and $\bigodot A$, there are five possibilities for the relative
positions of these two circles:
\begin{figure}[t!]
\vspace{-0.18cm}
\begin{center}
\includegraphics[width=0.578\textwidth]{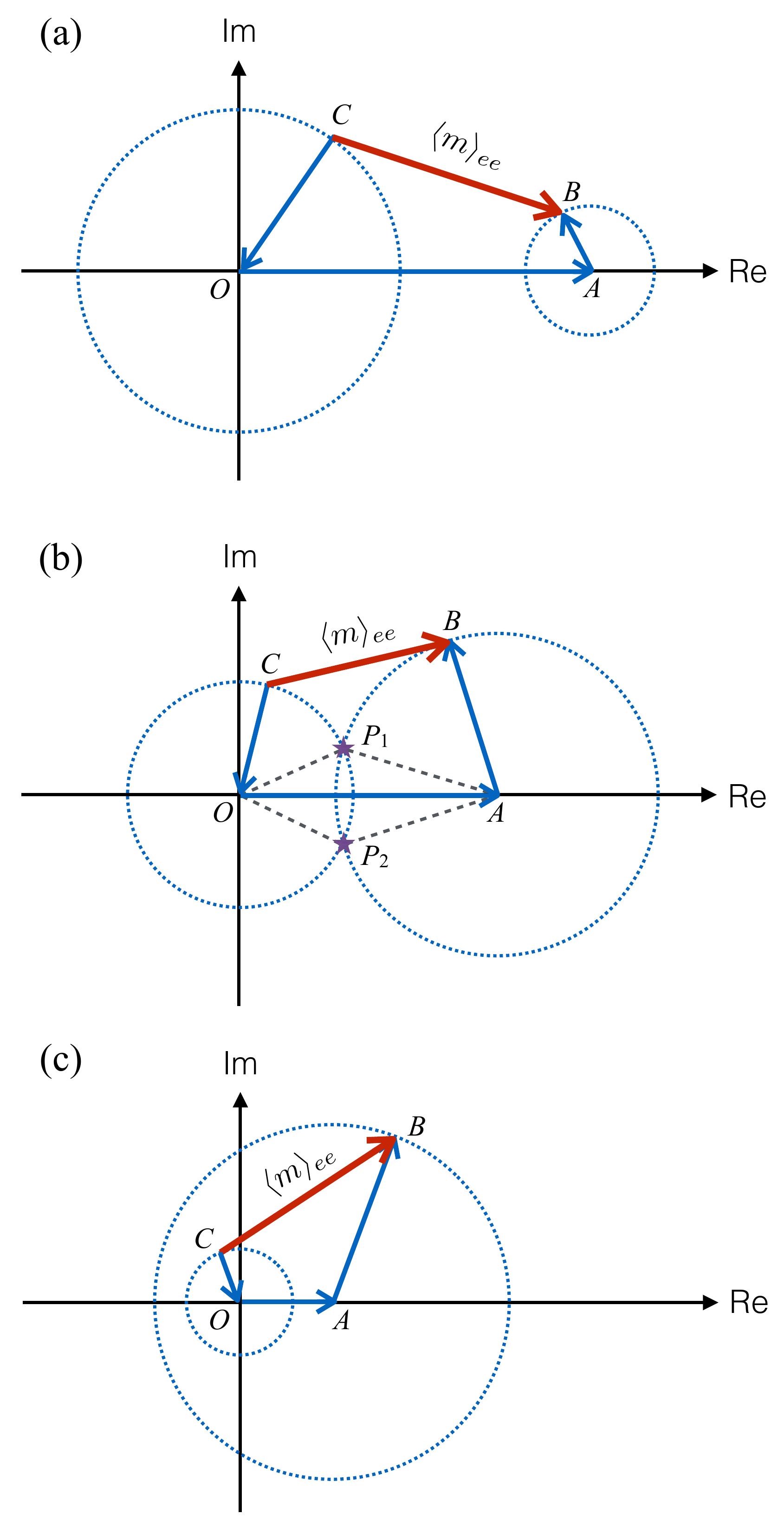}
\caption{The coupling-rod diagram of $\langle m\rangle^{}_{ee} \equiv
\protect\overrightarrow{CB}$ in the complex plane, where
$\protect\overrightarrow{OA} \equiv m^{}_2 |U^{}_{e2}|^2$,
$\protect\overrightarrow{AB} \equiv m^{}_1 |U^{}_{e1}|^2 e^{{\rm i}\rho}$ and
$\protect\overrightarrow{CO} \equiv m^{}_3 |U^{}_{e3}|^2 e^{{\rm i}\sigma}$.
If the neutrino mass ordering is normal, the three
configurations of $\langle m\rangle^{}_{ee}$ are all possible; but if
the neutrino mass ordering is inverted, then only Fig. 1(c) is allowed.}
\end{center}
\end{figure}
\begin{itemize}
\item     $AB + OC < OA$ as shown in Fig. 1(a),
or equivalently $m^{}_1 \cos^2\theta^{}_{12} + m^{}_3
\tan^2\theta^{}_{13} < m^{}_2 \sin^2\theta^{}_{12}$. Namely,
$\bigodot O$ and $\bigodot A$ are external to each other, and thus
$|\langle m\rangle^{}_{ee}| = BC >0$ holds. The allowed range of
$|\langle m\rangle^{}_{ee}|$ turns out to be $OA - AB - OC \leqslant
|\langle m\rangle^{}_{ee}| \leqslant OA + AB + OC$.

\item     $AB + OC = OA$, or equivalently $m^{}_1 \cos^2\theta^{}_{12}
+ m^{}_3 \tan^2\theta^{}_{13} = m^{}_2 \sin^2\theta^{}_{12}$.
Namely, $\bigodot O$ and $\bigodot A$ touch externally on the
horizontal axis. At the touching point $\rho = \sigma =\pi$ and
$|\langle m\rangle^{}_{ee}| = BC = 0$ hold. In this special case the
quadrilateral collapses into lines, and thus $m^{}_2$ can be
uniquely determined in terms of $\Delta m^2_{21}$, $\Delta
m^2_{31}$, $\theta^{}_{12}$ and $\theta^{}_{13}$. We find $m^{}_2
\simeq 8.4$ meV to $9.9$ meV by using the $3\sigma$ ranges of the
four input parameters \cite{Fogli}. It will be explained later on
that only the normal neutrino mass ordering is suitable for this
case.

\item     $|AB - OC| < OA < AB + OC$ as shown in Fig. 1(b),
where $\bigodot O$ and $\bigodot A$ intersect.
The two points of intersection imply
$|\langle m\rangle^{}_{ee}| = BC =0$; namely, the quadrilateral
collapses into a triangle. In this case, however, the two Majorana
phases should take some nontrivial values \cite{Xing2003}.

\item     $AB - OC = OA$, or equivalently
$m^{}_1 \cos^2\theta^{}_{12} - m^{}_3 \tan^2\theta^{}_{13} = m^{}_2
\sin^2\theta^{}_{12}$. Namely, $\bigodot O$ and $\bigodot A$ touch
internally on the horizontal axis. At the touching point $\rho =\pi$
and $\sigma =0$ hold, so does $|\langle m\rangle^{}_{ee}| = BC = 0$.
In this special case we find $m^{}_2 \simeq 9.5$ meV to $13.7$ meV
by inputting the $3\sigma$ ranges of the four parameters
\cite{Fogli}. Only the normal neutrino mass ordering is suitable for
this case.

\item     $AB - OC > OA$ as shown in Fig. 1(c), where  $\bigodot O$
and $\bigodot A$ do not touch and the former is contained in the
latter. In this case $|\langle m\rangle^{}_{ee}| = BC >0$ holds. The
allowed arrange of $|\langle m\rangle^{}_{ee}|$ turns out to be $AB
- OA - OC \leqslant |\langle m\rangle^{}_{ee}| \leqslant AB + OA +
OC$.
\end{itemize}
Note that the above discussions are not apparently subject to the
neutrino mass ordering, but the situation will be remarkably simpler
if the neutrino mass ordering is inverted. To see this point
clearly, let us take account of $|\Delta m^2_{31}| \sim 30 \Delta
m^2_{21}$ and $|U^{}_{e1}|^2 \sim 2 |U^{}_{e2}|^2 \sim 30
|U^{}_{e3}|^2$ as indicated by current experimental data
\cite{Fogli}. So $\Delta m^2_{31} <0$ leads us to $m^{}_3 < m^{}_1
\lesssim m^{}_2$, and the relative length of $OC$ becomes maximal
when the three neutrino masses are nearly degenerate (i.e., $m^{}_3
\lesssim m^{}_1 \lesssim m^{}_2$). In the latter case we are simply
left with $AB: OA: OC \sim |U^{}_{e1}|^2 : |U^{}_{e2}|^2:
|U^{}_{e3}|^2 \sim 30: 15: 1$, and thus it is impossible to satisfy
either $AB + OC \leqslant OA$ or $|AB - OC| \leqslant OA$. In other
words, only $AB - OC > OA$ can be satisfied in the inverted neutrino
mass ordering, and this observation keeps valid no matter whether
$m^{}_3$ is vanishing or close to the value of $m^{}_1$, or in
between. We arrive at two conclusions about $\langle
m\rangle^{}_{ee}$ in Fig. 1: (1) when $m^{}_1 < m^{}_2 < m^{}_3$
holds, the possibilities illustrated in Fig. 1(a), (b) and (c) are
all allowed, and they correspond to the values of $m^{}_1$ which are
small ($m^{}_1 \ll m^{}_2 \ll m^{}_3$), medium and large ($m^{}_1
\lesssim m^{}_2 \lesssim m^{}_3$), respectively; (2) when $m^{}_3 <
m^{}_1 < m^{}_2$ holds, only the possibility shown in Fig. 1(c) is
allowed, excluding $|\langle m\rangle^{}_{ee}| = 0$ in this case.

The geometric language has helped us to understand some salient
features of $\langle m\rangle^{}_{ee}$. We proceed to discuss the
maximum and minimum of $|\langle m\rangle^{}_{ee}|$ in an analytical
way. Eq. (3) can be rewritten as
\begin{eqnarray}
\langle m\rangle^{}_{ee} \hspace{-0.2cm} & = & \hspace{-0.2cm}
m^{}_2 |U^{}_{e2}|^2 \left[
1 + \frac{m^{}_1}{m^{}_2} \frac{|U^{}_{e1}|^2}{|U^{}_{e2}|^2}
e^{{\rm i}\rho} + \frac{m^{}_3}{m^{}_2} \frac{|U^{}_{e3}|^2}
{|U^{}_{e2}|^2} e^{{\rm i}\sigma} \right] \nonumber \\
\hspace{-0.2cm} & = & \hspace{-0.2cm} m^{}_2 \sin^2\theta^{}_{12}
\cos^2\theta^{}_{13} \left[ 1 + \sqrt{1 - \frac{\Delta
m^2_{21}}{m^2_2}} \cot^2\theta^{}_{12} e^{{\rm i}\rho} + \sqrt{1 -
\frac{\Delta m^2_{21}}{m^2_2} + \frac{\Delta m^2_{31}}{m^2_2}}
\frac{\displaystyle\tan^2\theta^{}_{13}}
{\displaystyle\sin^2\theta^{}_{12}} e^{{\rm i}\sigma} \right ] \; ,
\hspace{0.4cm}
\end{eqnarray}
where $m^{}_2 \geqslant \sqrt{\Delta m^2_{21}}$ must hold for the
normal mass ordering, or $m^{}_2 \geqslant \sqrt{\Delta m^2_{21} -
\Delta m^2_{31}}$ must hold for the inverted mass ordering. With the
help of the intuitive coupling-rod diagram of $\langle
m\rangle^{}_{ee}$ in Fig. 1, we can obtain the maximum or minimum of
$|\langle m\rangle^{}_{ee}|$ in two different cases:
\begin{itemize}
\item     $m^{}_1 < m^{}_2 < m^{}_3$.
In this case the maximum of $|\langle m\rangle^{}_{ee}| =BC$ can
be achieved in Fig. 1(a) when both $B$ and $C$ are located on the
horizontal axis and their distance is maximal (i.e., $\rho = \sigma
= 0$). Namely, $\left|\langle m\rangle^{}_{ee}\right|^{}_{\rm max} =
OA + AB + OC$, or equivalently
\begin{eqnarray}
\left|\langle m\rangle^{}_{ee}\right|^{}_{\rm max} = m^{}_2
\sin^2\theta^{}_{12} \cos^2\theta^{}_{13} \left[ 1 + \sqrt{1 -
\frac{\Delta m^2_{21}}{m^2_2}} \cot^2\theta^{}_{12} + \sqrt{1 -
\frac{\Delta m^2_{21}}{m^2_2} + \frac{\Delta m^2_{31}}{m^2_2}}
\frac{\displaystyle\tan^2\theta^{}_{13}}
{\displaystyle\sin^2\theta^{}_{12}} \right ] \; .
\end{eqnarray}
The minimum of $|\langle m\rangle^{}_{ee}|$ is a bit subtle as it
must arise from the maximal cancellation among its three complex
components \cite{Vissani}. Given $\Delta m^2_{31} >0$,
$\left|\langle m\rangle^{}_{ee}\right|^{}_{\rm min} = 0$ comes out
if $\bigodot O$ and $\bigodot A$ in Fig. 1 touch or intersect. When
$\bigodot O$ and $\bigodot A$ are external to each other as shown in
Fig. 1(a), $\left|\langle m\rangle^{}_{ee}\right|^{\rm (a)}_{\rm
min} = OA - AB - OC$, or equivalently
\begin{eqnarray}
\left|\langle m\rangle^{}_{ee}\right|^{\rm (a)}_{\rm min} = m^{}_2
\sin^2\theta^{}_{12} \cos^2\theta^{}_{13} \left[ 1 - \sqrt{1 -
\frac{\Delta m^2_{21}}{m^2_2}} \cot^2\theta^{}_{12} - \sqrt{1 -
\frac{\Delta m^2_{21}}{m^2_2} + \frac{\Delta m^2_{31}}{m^2_2}}
\frac{\displaystyle\tan^2\theta^{}_{13}}
{\displaystyle\sin^2\theta^{}_{12}} \right ] \; .
\end{eqnarray}
But when $\bigodot O$ is contained in $\bigodot A$ as shown in Fig.
1(c), $\left|\langle m\rangle^{}_{ee}\right|^{\rm (c)}_{\rm min} =
AB - OA - OC$; namely,
\begin{eqnarray}
\left|\langle m\rangle^{}_{ee}\right|^{\rm (c)}_{\rm min} =
m^{}_2 \sin^2\theta^{}_{12} \cos^2\theta^{}_{13} \left[
\sqrt{1 - \frac{\Delta m^2_{21}}{m^2_2}}
\cot^2\theta^{}_{12} - 1 - \sqrt{1 - \frac{\Delta m^2_{21}}{m^2_2}
+ \frac{\Delta m^2_{31}}{m^2_2}} \frac{\displaystyle\tan^2\theta^{}_{13}}
{\displaystyle\sin^2\theta^{}_{12}} \right ] \; .
\end{eqnarray}

\item     $m^{}_3 < m^{}_1 < m^{}_2$.
In this case $\langle m\rangle^{}_{ee}$ is uniquely described by
Fig. 1(c), and its maximum or minimum can be obtained when both $B$
and $C$ are located on the horizontal axis and their distance is
maximal (i.e., $\rho = \sigma =0$) or minimal (i.e., $\rho =\pi$ and
$\sigma =0$). The expressions of $\left|\langle
m\rangle^{}_{ee}\right|^{}_{\rm max}$ and $\left|\langle
m\rangle^{}_{ee}\right|^{}_{\rm min}$ are the same as Eqs. (7) and
(9), but the sign of $\Delta m^2_{31}$ is now negative.
\end{itemize}
We plot the dependence of $|\langle m\rangle^{}_{ee}|$ on $m^{}_2$
in Fig. 2 by inputting the $3\sigma$ ranges of $\Delta m^2_{21}$,
$\Delta m^2_{31}$, $\theta^{}_{12}$ and $\theta^{}_{13}$
\cite{Fogli} and allowing the relevant CP-violating phases to vary
between $0$ and $2\pi$. The numerical results are consistent with
the above analytical observation. In particular, the upper or lower
bound of $|\langle m\rangle^{}_{ee}|$ in the inverted neutrino mass
ordering follows almost the same behavior as that in the normal
case, because both of them are governed by Eq. (7) or Eq. (9) with
$\Delta m^2_{31}$ taking the opposite signs. Thanks to $m^{}_2
\geqslant \sqrt{\Delta m^2_{21}}$ in the normal case, the allowed
region of $|\langle m\rangle^{}_{ee}|$ looks like a hockey stick.
But it has to be cut shorter by $|\langle m\rangle^{}_{ee}| <
0.19$---$0.45$ eV (or $0.2$---$0.4$ eV) set by the EXO-200
\cite{EXO} (or GERDA \cite{GERDA}) experiment at the $90\%$
confidence level, and by $m^{}_2 < 0.08$ eV that is derived from the
cosmological constraint $m^{}_1 + m^{}_2 + m^{}_3 < 0.23$ eV set by
the Planck data \cite{Planck} at the $95\%$ confidence level.
\begin{figure}[t!]
\vspace{-0.5cm}
\begin{center}
\includegraphics[width=0.63\textwidth]{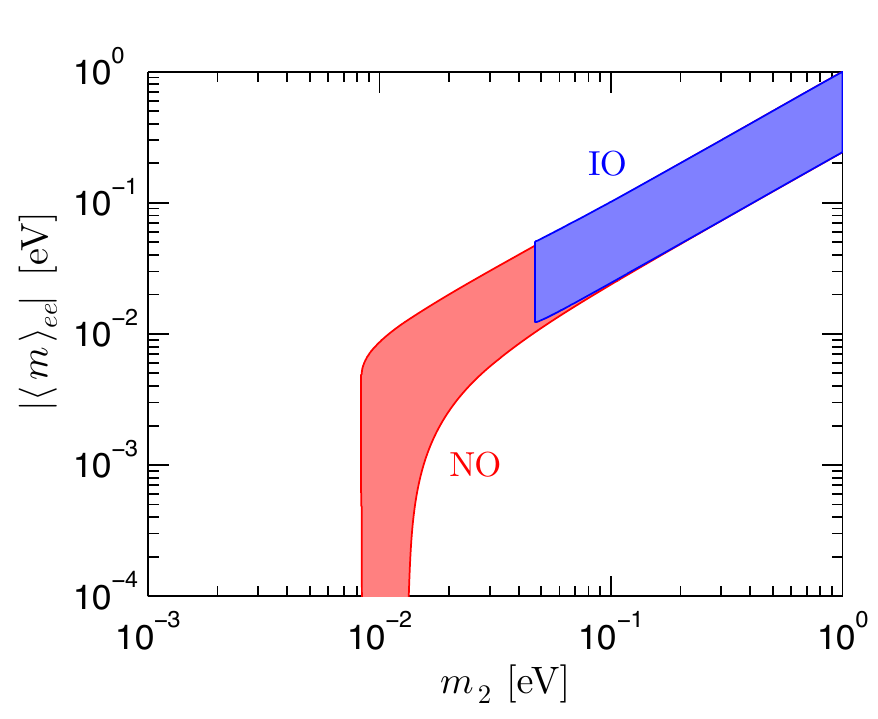}
\vspace{-0.1cm} \caption{The dependence of $\langle
m\rangle^{}_{ee}$ on $m^{}_2$ in the normal ordering (NO) or
inverted ordering (IO) of three neutrino masses, where the $3\sigma$
ranges of $\Delta m^2_{21}$, $\Delta m^2_{31}$, $\theta^{}_{12}$ and
$\theta^{}_{13}$ \cite{Fogli} have been input, and the relevant
CP-violating phases are allowed to vary between $0$ and $2\pi$.}
\end{center}
\end{figure}

On the other hand, Fig. 2 tells us that $|\langle m\rangle^{}_{ee}|$
tends to approach a very small value and even vanish when $m^{}_2$
is not far away from its lower bound $\sqrt{\Delta m^2_{21}} \simeq
8.7 \times 10^{-3} ~{\rm eV}$ (i.e., when the base vector
$\overrightarrow{OA}$ in Fig. 1 is roughly as short as possible).
This observation is certainly true, as geometrically shown in Fig.
1, where a sufficiently short $OA$ means that $\bigodot O$ and
$\bigodot A$ are more likely to touch or intersect, allowing
$|\langle m\rangle^{}_{ee}| \to 0$ to naturally take place.

In case the experimental sensitivity has been good enough but a
signal of the $0\nu\beta\beta$ decay remains absent, there might be
three possibilities: (1) massive neutrinos are the Dirac particles;
(2) $|\langle m\rangle^{}_{ee}|$ itself is too small; (3) new
physics corrects $\langle m\rangle^{}_{ee}$ in a destructive way to
make its size too small. Although $|\langle m\rangle^{}_{ee}| \to 0$
implies that it will be impossible to identify the Majorana nature
of massive neutrinos via the $0\nu\beta\beta$ decay, this special
case is interesting in the sense that it allows us to determine the
two Majorana CP-violating phases \cite{Xing2003}. As one can see in
Fig. 1(b), the quadrilateral becomes a triangle (i.e., $\triangle
OP^{}_1 A$ or $\triangle OP^{}_2 A$) in the $|\langle
m\rangle^{}_{ee}| = BC \to 0$ limit. Because of $\triangle OP^{}_2 A
\cong \triangle OP^{}_1 A$, we have $\rho = \pi \mp \angle OAP^{}_1$
and $\sigma = \pi \pm \angle AOP^{}_1$ for $\triangle OP^{}_1 A$ and
$\triangle OP^{}_2 A$, respectively. In this case the two inner
angles of $\triangle OP^{}_1 A$ can be calculated from its three
sides $OA$, $AP^{}_1 = AB$ and $OP^{}_1 = OC$ by means of the cosine
theorem. An analytical discussion has been done in Ref.
\cite{Xing2003} by taking $\langle m\rangle^{}_{ee} =0$ in Eq. (6)
to obtain two constraint equations for $\rho$ and $\sigma$. From a
phenomenological point of view, however, one certainly prefers that
the neutrino masses are nearly degenerate or have the inverted
ordering, so as to assure $|\langle m\rangle^{}_{ee}| \geq 10$ meV
which should be accessible in the next-generation $0\nu\beta\beta$
experiments.

Once $|\langle m\rangle^{}_{ee}|$ is determined from a measurement
of the CP-conserving $0\nu\beta\beta$ decay, one will be able to
partly constrain the absolute neutrino mass scale and two Majorana
CP-violating phases \cite{Rode}. There are two special cases, in
which $m^{}_2$ can be fixed and $\langle m\rangle^{}_{ee}$ only
involves a single phase parameter:
\begin{itemize}
\item     $m^{}_1 =0$, which leads to $AB =0$. In this case
$\bigodot A$ shrinks into a point, and thus the quadrilateral in
Fig. 1 is simplified to $\triangle OAC$. As a result, $|\langle
m\rangle^{}_{ee}|$ only depends on a single CP-violating phase:
\begin{eqnarray}
|\langle m\rangle^{}_{ee}| = \sqrt{\Delta m^2_{21} s^4_{12} c^4_{13}
+ \Delta m^2_{31} s^4_{13} + 2\sqrt{\Delta m^2_{21} \Delta m^2_{31}}
\ c^2_{12} s^2_{12} s^2_{13} \cos\sigma} \; .
\end{eqnarray}

\item     $m^{}_3 =0$, which leads to $OC =0$. In this case
the quadrilateral in Fig. 1 is simplified to $\triangle OAB$, and
the magnitude of $\langle m\rangle^{}_{ee}$ turns out to be
\begin{eqnarray}
|\langle m\rangle^{}_{ee}| = c^2_{13} \sqrt{\left(\Delta m^2_{21}
-\Delta m^2_{31}\right) s^4_{12} - \Delta m^2_{31} c^4_{12} + 2
\sqrt{\Delta m^2_{31} \left(\Delta m^2_{31} - \Delta m^2_{21}
\right)} \ c^2_{12} s^2_{12} \cos\rho} \; ,
\end{eqnarray}
which is also dependent upon a single CP-violating phase.
\end{itemize}
In either case the range of $|\langle m\rangle^{}_{ee}|$ can easily
be determined by allowing the respective CP-violating phase to vary
from 0 to $2\pi$. With the help of the $3\sigma$ ranges of $\Delta
m^2_{21}$, $\Delta m^2_{31}$, $\theta^{}_{12}$ and $\theta^{}_{13}$
\cite{Fogli}, we immediately arrive at $0.68$ meV $\leqslant
|\langle m\rangle^{}_{ee}| \leqslant 4.7$ meV in the $m^{}_1 =0$
case, and $12.4$ meV $\leqslant |\langle m\rangle^{}_{ee}| \leqslant
50.1$ meV in the $m^{}_3 =0$ case. The latter is of course more
promising in the future $0\nu\beta\beta$ experiments.

\section{Comments on $\langle m\rangle^{}_{\alpha\beta}$ and new physics}

The coupling-rod diagram of $\langle m\rangle^{}_{ee}$ in Fig. 1 can
be extended to geometrically describe $\langle
m\rangle^{}_{\alpha\beta}$ (for $\alpha, \beta = e, \mu, \tau$) in
general. For each individual $\langle m\rangle^{}_{\alpha\beta}$, it
is always possible to adopt a proper parametrization and phase
convention of $U$ to make $m^{}_2 U^{}_{\alpha 2} U^{}_{\beta 2}$
real and positive. A typical example of this kind is \cite{FX}
\footnote{Among the nine possible parametrizations of $U$ listed in
Ref. \cite{FX}, Patterns (5), (6) and (7) satisfy the requirement
because the relevant $U^{}_{\alpha 2}$ elements (for $\alpha = e,
\mu, \tau$) are all independent of the ``Dirac" CP-violating phase.
Hence these elements can also be arranged to be independent of the
two Majorana CP-violating phases in a very straightforward way.}
\begin{eqnarray}
U = \left( \begin{matrix} c^{\prime}_{12} c^{\prime}_{13} &
s^{\prime}_{12} & -c^{\prime}_{12} s^{\prime}_{13} \cr -
c^{\prime}_{12} s^{\prime}_{12} c^{\prime}_{13} + s^{\prime}_{12}
s^{\prime}_{13} e^{-{\rm i} \delta^\prime} & c^{\prime}_{12}
c^{\prime}_{23} & s^{\prime}_{12} s^{\prime}_{13} c^{\prime}_{23} +
c^{\prime}_{13} s^{\prime}_{23} e^{-{\rm i} \delta^\prime} \cr
-s^{\prime}_{12} c^{\prime}_{13} s^{\prime}_{23} - s^{\prime}_{13}
c^{\prime}_{23} e^{-{\rm i} \delta^\prime} & c^{\prime}_{12}
s^{\prime}_{23} & s^{\prime}_{12} s^{\prime}_{13} s^{\prime}_{23} -
c^{\prime}_{13} c^{\prime}_{23} e^{-{\rm i} \delta^\prime} \cr
\end{matrix} \right) P^{\prime}_\nu \; ,
\end{eqnarray}
where $c^{\prime}_{ij} \equiv \cos\theta^{\prime}_{ij}$,
$s^{\prime}_{ij} \equiv \sin\theta^{\prime}_{ij}$ (for $ij = 12, 13,
23$), and $P^{\prime}_\nu$ is a diagonal matrix containing the other
two independent CP-violating phases. Its connection to the standard
parametrization of $U$ in Eq. (2) is straightforward. In this case
one may express $\langle m\rangle^{}_{\alpha\beta}$ as a sum of
three vectors in the complex plane:
\begin{eqnarray}
\langle m\rangle^{}_{\alpha\beta} \equiv \overrightarrow{CB} =
\overrightarrow{O A} +  \overrightarrow{AB} +  \overrightarrow{C O}
\; ,
\end{eqnarray}
where
\begin{eqnarray}
\overrightarrow{OA} \hspace{-0.2cm} & \equiv & \hspace{-0.2cm}
m^{}_2 U^{}_{\alpha 2} U^{}_{\beta 2} = m^{}_2
|U^{}_{\alpha 2} U^{}_{\beta 2}| \; ,
\nonumber \\
\overrightarrow{AB} \hspace{-0.2cm} & \equiv & \hspace{-0.2cm}
m^{}_1 U^{}_{\alpha 1} U^{}_{\beta 1} = m^{}_1 |U^{}_{\alpha 1}
U^{}_{\beta 1}| e^{{\rm i}\rho^\prime} \; ,
\nonumber \\
\overrightarrow{CO} \hspace{-0.2cm} & \equiv & \hspace{-0.2cm}
m^{}_3 U^{}_{\alpha 3} U^{}_{\beta 3} = m^{}_3 |U^{}_{\alpha 3}
U^{}_{\beta 3}| e^{{\rm i}\sigma^\prime} \; . \hspace{0.1cm}
\end{eqnarray}
Note that the phase parameters $\rho^\prime$ and $\sigma^\prime$
depend on the subscripts $\alpha$ and $\beta$. Of course, Eq. (13)
stands for a quadrilateral which is quite similar to the
coupling-rod diagram of $\langle m\rangle^{}_{ee}$ in Fig. 1.
Depending on the radii of $\bigodot O$ and $\bigodot A$ with respect
to $\langle m\rangle^{}_{\alpha\beta}$, Fig. 1(a), (b) and (c) may
analogously describe the relative positions of these two circles. An
exceptional case, which is not shown by Fig. 1, is that $\bigodot A$
is likely to be contained in $\bigodot O$ (i.e., $OC > OA + AB$) for
some of the effective Majorana mass terms
\footnote{If the standard parametrization of $U$ in Eq. (2) is
applied to Eqs. (13) and (14), a nontrivial issue associated with
$\langle m\rangle^{}_{\alpha\beta}$ (for $\alpha\beta \neq ee$) will
be that $OA$ and $AB$ become dependent upon the CP-violating phase
$\delta$. The latter remains unknown, and thus its uncertainty will
more or less complicate our discussions.}.

While we do not go into details of the coupling-rod diagrams of
$\langle m\rangle^{}_{\alpha\beta}$ in the present work, it is
desirable for us to stress the importance of probing these effective
neutrino masses in all the possible lepton-number-violating
processes (e.g., $\langle m\rangle^{}_{\alpha\beta}$ can play an
important role in the probabilities of neutrino-antineutrino
oscillations and in the rates of $H^{++}\to \ell^+_\alpha
\ell^+_\beta$ and $H^+\to \ell^{}_\alpha \overline{\nu}$ decays
\cite{XZ2013}).

Next, let us briefly comment on possible corrections to $\langle
m\rangle^{}_{\alpha\beta}$ from underlying new physics. For
simplicity, we assume that the effect of new physics is not
correlated with the standard three-flavor $\langle
m\rangle^{}_{\alpha\beta}$ in the leading-order approximation but
only provides a linear correction to $\langle
m\rangle^{}_{\alpha\beta}$ in the form of
\begin{eqnarray}
\langle m\rangle^\prime_{\alpha\beta} = \langle
m\rangle^{}_{\alpha\beta} + {\rm new ~ physics} \; .
\end{eqnarray}
The source of new physics is unknown to us, but some typical
examples like the sterile neutrinos and the R-parity-violating
supersymmetry have been explored in the literature \cite{Rode}.
Because the new-physics term generally involves one or more
CP-violating phases, a potential cancellation between it and
$\langle m\rangle^{}_{\alpha\beta}$ is likely to lead to vanishing
or vanishingly small $\langle m\rangle^{\prime}_{\alpha\beta}$
\cite{Li}. To illustrate, Fig. 3 shows a coupling-rod-like diagram
of $\langle m\rangle^{\prime}_{ee}$ in the presence of new physics
described by the $\overrightarrow{BD}$ vector:
\begin{eqnarray}
\langle m\rangle^\prime_{\alpha\beta} \equiv \overrightarrow{CD} =
\langle m\rangle^{}_{\alpha\beta} + \overrightarrow{BD} =
\overrightarrow{C O} + \overrightarrow{O A} +  \overrightarrow{AB} +
\overrightarrow{BD} \; .
\end{eqnarray}
Depending on the size and phase of $\overrightarrow{BD}$, a number
of different configurations of the vectors in Eq. (16) are possible.
Fig. 3 only illustrates two simple cases: (A) $\bigodot O$ and
$\bigodot A$ are external to each other; and (B) $\bigodot O$ is
contained in $\bigodot A$. Once the nature of new physics is
quantitatively fixed, one may give a detailed geometrical
description of the salient features of $\langle
m\rangle^{\prime}_{ee}$ as what we have done for $\langle
m\rangle^{}_{ee}$ itself.
\begin{figure}[t!]
\vspace{-0.18cm}
\begin{center}
\includegraphics[width=0.606\textwidth]{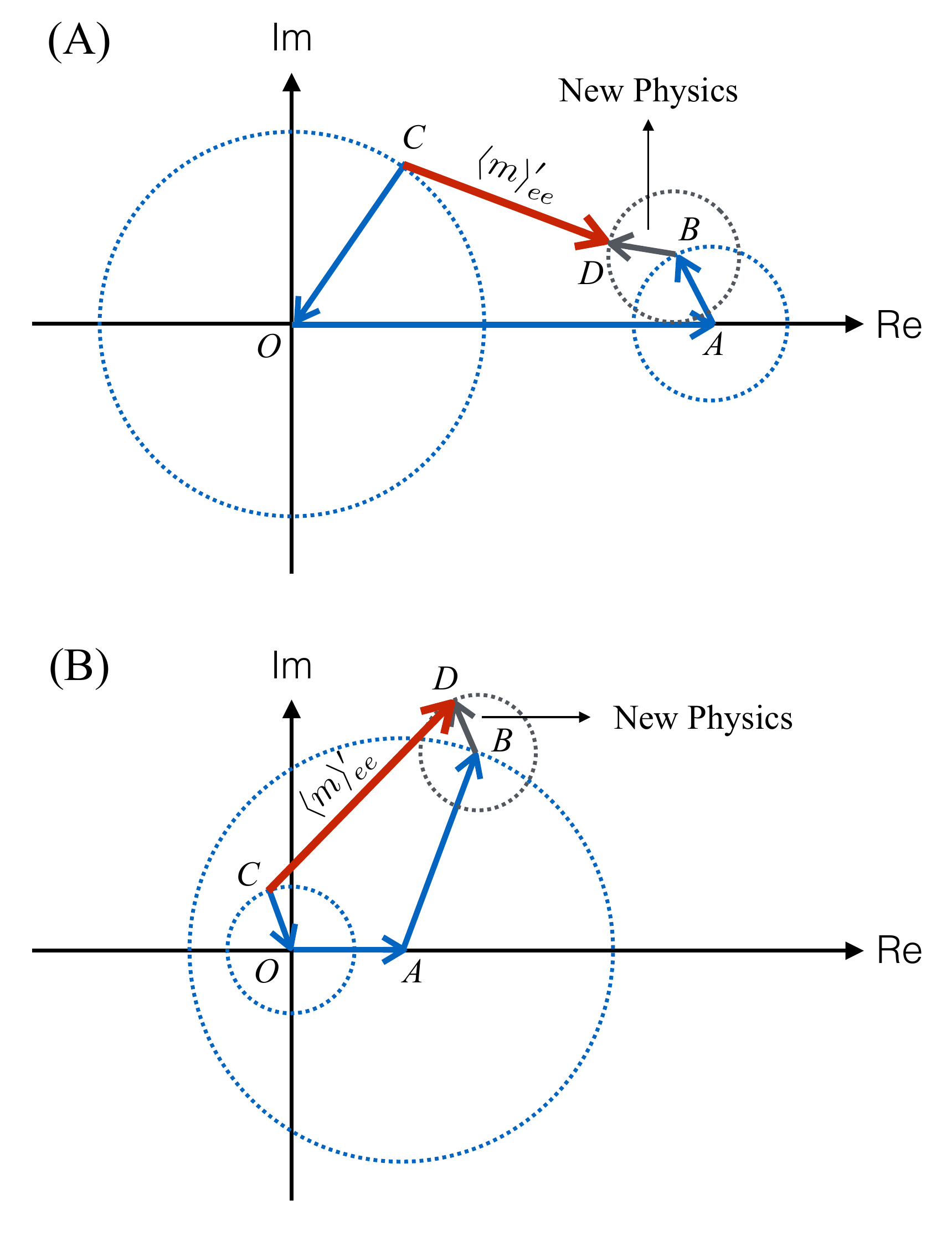}
\caption{A coupling-rod-like diagram of $\langle
m\rangle^{\prime}_{ee} \equiv \protect\overrightarrow{CD}$ in the
complex plane: (A) $\bigodot O$ and $\bigodot A$ are external to
each other; and (B) $\bigodot O$ is contained in $\bigodot A$, where
$\protect\overrightarrow{OA} \equiv m^{}_2 |U^{}_{e2}|^2$,
$\protect\overrightarrow{AB} \equiv m^{}_1 |U^{}_{e1}|^2 e^{{\rm
i}\rho}$, $\protect\overrightarrow{CO} \equiv m^{}_3 |U^{}_{e3}|^2
e^{{\rm i}\sigma}$, and the new physics contribution is described by
the vector $\protect\overrightarrow{BD}$ for illustration.}
\end{center}
\end{figure}

\section{Summary}

The $0\nu\beta\beta$ decay has long been  recognized as a unique
process to identify the Majorana nature of massive neutrinos, and
hence searching for its signal becomes almost the most important
task in the non-oscillation aspects of today's neutrino physics. Now
that the $0\nu\beta\beta$ decay rate depends on the size of the
effective Majorana neutrino mass $\langle m\rangle^{}_{ee}$, it is
desirable to explore the salient features of $\langle
m\rangle^{}_{ee}$ in a phenomenologically favored way. In this work
we have put forward a novel coupling-rod diagram to describe
$\langle m\rangle^{}_{ee}$ in the complex plane, by which the
effects of the neutrino mass ordering and CP-violating phases on
$\langle m\rangle^{}_{ee}$ can be intuitively understood. We have
shown that this simple geometric language allows us to easily obtain
the maximum and minimum of $|\langle m\rangle^{}_{ee}|$. It remains
usable even if there is a kind of new physics contributing to
$\langle m\rangle^{}_{ee}$. It can also be extended to describe the
effective Majorana masses $\langle m\rangle^{}_{e\mu}$, $\langle
m\rangle^{}_{e\tau}$, $\langle m\rangle^{}_{\mu\mu}$, $\langle
m\rangle^{}_{\mu\tau}$ and $\langle m\rangle^{}_{\tau\tau}$ which
may appear in some other lepton-number-violating processes, if a
proper parametrization and phase convention of the lepton mixing
matrix $U$ is adopted.

Although the geometrical and analytical descriptions of $\langle
m\rangle^{}_{ee}$ are ``scientifically indistinguishable", ``they
are not psychologically identical" in making the underlying physics
more transparent \cite{Feynman}. For this reason we expect that the
coupling-rod diagram of $\langle m\rangle^{}_{ee}$, just like the
unitarity triangles of quark and lepton flavor mixing matrices, can
prove to be useful in neutrino phenomenology.

\vspace{0.5cm}

We are indebted to W.L. Guo, Y.F. Li, L.J. Wen and S. Zhou for
helpful discussions. This work was supported in part by the National
Natural Science Foundation of China under grant No. 11135009.

\end{document}